\documentclass[twocolumn,showpacs,aps,prl,superscriptaddress,floatfix,amsmath,amssymb,nofootinbib]{revtex4}

\usepackage{graphicx}
\usepackage{lineno}

\input pubboard/babarsym

\long\def\inst#1{\par\nobreak\kern 4pt\nobreak
    {\it #1}\par\vskip 10pt plus 3pt minus 3pt}

\begin{document}

\begin{flushleft}
\babar-PUB-15/007\\
SLAC-PUB-16337
\end{flushleft}

\title{\large\bf\boldmath
Observation of $\Bb \to D^{(*)} \pip\pim \ell^-\nub$ decays in $\epem$ collisions at the 
$\Upsilon(4S)$ resonance}

%
\author{J.~P.~Lees}
\author{V.~Poireau}
\author{V.~Tisserand}
\affiliation{Laboratoire d'Annecy-le-Vieux de Physique des Particules (LAPP), Universit\'e de Savoie, CNRS/IN2P3,  F-74941 Annecy-Le-Vieux, France}
\author{E.~Grauges}
\affiliation{Universitat de Barcelona, Facultat de Fisica, Departament ECM, E-08028 Barcelona, Spain }
\author{A.~Palano$^{ab}$ }
\affiliation{INFN Sezione di Bari$^{a}$; Dipartimento di Fisica, Universit\`a di Bari$^{b}$, I-70126 Bari, Italy }
\author{G.~Eigen}
\author{B.~Stugu}
\affiliation{University of Bergen, Institute of Physics, N-5007 Bergen, Norway }
\author{D.~N.~Brown}
\author{L.~T.~Kerth}
\author{Yu.~G.~Kolomensky}
\author{M.~J.~Lee}
\author{G.~Lynch}
\affiliation{Lawrence Berkeley National Laboratory and University of California, Berkeley, California 94720, USA }
% ...
\collaboration{The \babar\ Collaboration}
\noaffiliation

\setcounter{footnote}{0}

\begin{abstract}

  We report on measurements of the decays of $\Bb$ mesons into the semileptonic final states 
  $\Bb \to D^{(*)} \pip\pim \ell^-\nub$, where $D^{(*)}$ represents a $D$ or $\Dstar$ meson
  and $\ell^-$ is an electron or a muon.
  These measurements are based on $471\times 10^6$ \BB\ 
  pairs recorded 
  with the \babar\ detector at the SLAC \abf\ \pep2. 
  We determine the branching fraction ratios 
  $R^{(*)}_{\pip\pim} = \mathcal{B}(\Bb \to D^{(*)} \pip\pim \ell^- \nub)/\mathcal{B}(\Bb \to D^{(*)} \ell^-\nub)$
  using events in which the second $B$
  meson is fully reconstructed.
We find
  $R_{\pip\pim} = 0.067 \pm 0.010 \pm 0.008$ and
  $R^*_{\pip\pim} = 0.019 \pm 0.005 \pm 0.004$,   
where the first uncertainty is statistical and the second is systematic.  Based on these results,
we estimate that $\Bb\to D^{(*)}\pi\pi\ell^-\nub$ decays, where $\pi$ denotes both a $\pi^\pm$ and $\pi^0$ meson, account for up to half the difference between the 
measured inclusive semileptonic branching fraction to charm hadrons and the corresponding
sum of previously measured exclusive branching fractions.
\end{abstract}


\maketitle

The semileptonic decays of $B$ mesons to final states containing a charm quark allow a 
measurement of the magnitude of the CKM matrix element~\cite{ref:C,ref:KM}  
$\Vcb$, a fundamental parameter
in the standard model (SM) of particle physics
that plays an important role in unitarity tests sensitive to physics beyond 
the SM~\cite{ref:NewPhysics}.  Determinations 
of $\Vcb$ from inclusive semileptonic decays $\Bb\to (X_c)\ell^-\nub,$\footnote{Throughout this 
Letter, whenever a decay mode is given, the charge conjugate is also implied.  
} where 
the hadronic state $X_c$ is not reconstructed,
and those from the exclusive semileptonic decays
$\Bb\to D^*\ell^-\nub$ and $\Bb\to D\ell^-\nub$,
differ by
nearly three standard deviations ($3\sigma$), as discussed on page 1208 of Ref.~\cite{ref:PDG2014}.
The measured exclusive $\Bb\to X_c\ell^-\nub$ decays, 
$\Bb\to D^{(*)}\ell^-\nub$,\footnote{The 
notation $D^{(*)}$ denotes \Dz, \Dp, \Dstarz\ and \Dstarp\ mesons.
}
$\Bb\to D^{(*)}\pi\ell^-\nub$, and $\Bb\to D_s^{(*)+}\Km\ell^-\nub$~\cite{ref:PDG2014},
account for 
only $85\pm 2\% $~\cite{ref:BLT2012} of the inclusive rate for 
semileptonic $\Bb$ decays to charm final states.
The decay modes measured in this Letter account for part of this difference.  
They also provide experimental
information needed to quantify background-related systematic uncertainties in measurements 
of $\Bb\to D^{(*)}\tau\nub$ decays, which are sensitive to new physics contributions.  A 
measurement~\cite{ref:DtaunuBaBar} of these decays shows a $3.4\sigma$ 
deviation from the SM, and independent measurements~\cite{ref:DtaunuBelle,ref:DtaunuLHCb}
also exceed SM expectations.

We search for semileptonic decays of a $B$ meson to a $D$ or $D^*$ meson and
two additional charged pions, and measure branching fraction ratios 
$R^{(*)}_{\pip\pim} = \mathcal{B}(\Bb \to D^{(*)} \pip\pim \ell^- \nub)/\mathcal{B}(\Bb \to D^{(*)} \ell^-\nub)$
relative to the topologically similar decays $\Bb\to D^{(*)}\ell^-\nub$.
The results are based on the complete sample of $\epem$
collision data collected at the $\Upsilon(4S)$ resonance with the \babar\ detector
at the SLAC \pep2\ storage ring, corresponding to $471\times 10^6$ \BB\ decays ($454\,$fb$^{-1}$~\cite{ref:Lumipreprint}).
An additional $40$~fb$^{-1}$ sample,
collected at center-of-mass (CM) energies just below the \BB\ 
threshold,
is used to verify the modeling of background from 
$\epem\to f\bar{f}(\gamma)$ continuum processes with $f = u, d, s, c, \tau$.

The \babar\ detector, as well as the reconstruction and particle identification algorithms, are described in detail elsewhere~\cite{ref:BaBarDetector}. 
The analysis uses Monte Carlo (MC) simulated event 
samples to determine efficiencies and to model backgrounds.
Simulated $\BB$ decays are produced with the EvtGen~\cite{ref:EvtGen} generator, with final-state
radiation described using the PHOTOS~\cite{ref:PHOTOS} program.
Continuum $\epem\to q\bar{q}$ events are generated with the JETSET~\cite{ref:JETSET} program,
and $\epem\to \taup\taum$ events with the KK2F~\cite{ref:KK2F} program.
  The world averages quoted in Ref.~\cite{ref:PDG2014} are used for branching fractions 
  and form factor parameters.
  The GEANT4~\cite{ref:GEANT4} package is used to model the detector and detector response.

  The intermediate process through which $D^{(*)}\pip\pim$ states arise in semileptonic $B$ decays
  is unknown.
  We consider production via 
(1) 3-body phase-space decays, $X_c\to D^{(*)}\pi\pi$,
(2) $X_c\to D^{(*)}f_0(500)$ decays with $f_0(500)\to\pi\pi$, 
(3) sequential decays $X_c\to Y_c\pi$ followed by $Y_c\to D^{(*)}\pi$, and 
(4) $X_c\to D^{(*)}\rho$ decays with $\rho\to\pi\pi$,
  where $X_c$ is one of $D_1(2420)$, $D(2S)$, or $D^*(2S)$, and $Y_c$ is one of $D_1(2430)$, $D_0^*$, 
  or $D_2^*$. The $D^{(*)}(2S)$ states are the first radial excitations of the ground state 
  $D^{(*)}$ mesons, and are modeled as in Ref.~\cite{ref:BLT2012}.  
  Our nominal signal model consists of 3-body phase-space 
  $X_c\to D^{(*)}\pi\pi$ decays with an equal mix of $X_c$ mesons.

  We reconstruct events of the type $\epem \to \FourS \to \BB$. One of the \B\ mesons ($B_{\rm tag}$) is fully 
  reconstructed in a hadronic final state.
  To reconstruct a $B_{\rm tag}$ candidate, a
  seed (one of $D^{(*)}, D_s^{(*)+}$, or $\jpsi$) is combined
  with up to five additional particles (pions and/or kaons), as described in Ref.~\cite{ref:DtaunuBaBar}.
  The $B_{\rm tag}$ candidates are required to have an energy-substituted mass 
  $m_{ES} \equiv \sqrt{ s/4 - \vert \vec{p}_{\rm tag}\vert^2 } > 5.27 \gevcc$, 
  and a difference between the 
  beam energy and the reconstructed energy of the $B_{\rm tag}$ candidate
  $|\Delta E|\equiv |E_{\rm tag}-\sqrt{s}/2| \leq 0.09 \gev$, 
  where $\sqrt{s}$ is the total $\epem$ energy and $\vec{p}_{\rm tag}$ and $E_{\rm tag}$ are the 
  measured $B_{\rm tag}$ momentum and energy in the \epem CM frame. 

  For each $B_{\rm tag}$ candidate, we use the remaining particles in the event to search for
  signal \Bb\ meson candidates involving a $D$ or $D^*$ meson, a charged lepton, and up to two
  charged pions.
  The $\Dz$ and $\Dp$ candidates are reconstructed in final states involving up to four charged
  pions or kaons, up to one $K^0_S\to\pip\pim$ decay, and up to one $\piz\to\gamma\gamma$ decay.  
  We require $1.845 < m(D^+) < 1.895 \gevcc$ and $1.840 < m(\Dz) < 1.890 \gevcc$.  
  The $\Dstar$ mesons are reconstructed in 
  $\Dstarz\to \Dz\piz$, $\Dstarz\to \Dz\gamma$, $\Dstarp\to \Dz\pip$, and $\Dstarp\to \Dp\piz$ decays. 
  Electrons and muons are identified using multivariate 
  techniques based on information from the tracking detectors, calorimeter, and muon system, 
  and are required to have a momentum larger than
  $0.6 \gevc$ in the CM frame.
  We reject electrons consistent with photon conversions and Dalitz decays of $\piz$ mesons.
  In cases where the flavor of the $D^{(*)}$ meson is determined by its decay products,
  only combinations with the correct $D^{(*)}\ell^-$ charge-flavor correlation are retained. 
  For each $B_{\rm tag}D^{(*)}\ell^-$ candidate we allow up to two additional charged tracks in the event, 
  resulting in a sample consisting of $B_{\rm tag}D^{(*)} (n \pi) \ell^-$ candidates, with ``signal
  pion" multiplicity $n=0$, $1$ or $2$.  Our measurement is based on the $n=0$ and $n=2$ samples,
  while the $n=1$ sample is used to reject backgrounds in the $n=2$ sample.
  
  Only candidates for which all charged tracks are assigned to one or the 
  other $B$ meson, and where the net charge of the event is zero, are
  considered further.
  Charged $B_{\rm tag}$ candidates are required to have charge opposite that of the lepton candidate.
  We calculate $E_{\rm extra}$, the energy sum of all calorimeter energy 
  clusters with energy greater than $80 \mev$ that are not used in the reconstruction of the
  $B$ candidates, and require $E_{\rm extra} \leq 0.4 \gev$. 
  After these criteria are applied, the remaining events have on average about two 
  $\FourS\to\B_{\rm tag}\Bb$ candidates per signal channel.  The candidate in each $D^{(*)}(n\pi)\ell^-$
  channel with the smallest $\vert \Delta E \vert$ is retained.
  
  Each $\FourS\to B_{\rm tag}\Bb$ candidate is fit to the hypothesized decay topology, imposing vertex 
  and mass constraints on intermediate states in order to improve the resolution.
  The four-momentum of the $B_{\rm tag}D^{(*)}(n\pi)\ell^-$ candidate
  is subtracted from that of the initial $\epem$ state to determine the 
  four-momentum $p_{\rm miss}=(E_{\rm miss},\vec{p}_{\rm miss})$.
  For events in which a single neutrino is the only missing particle, 
  the difference $U\equiv E_{\rm miss}-\vert \vec{p}_{\rm miss}\vert c$ peaks at zero with a 
  resolution of $\approx 0.1\gev$; $U$ is used to discriminate against events with 
  additional missing particles.  
  In contrast to the commonly used missing-mass-squared,
  which is proportional to $E_{\rm miss}+|\vec{p}_{\rm miss}|\approx 2E_{\rm miss}$,
  $U$ does not depend 
  directly on the modeling of $E_{\rm miss}$ and thus on the decay dynamics.
  Hadronic $B$ decays for which all final-state particles are reconstructed, and
  in which a hadron is misidentified as an electron or muon, have 
  $E_{\rm miss}\approx |\vec{p}_{\rm miss}|\approx 0$: 
  we require $\vert \vec{p}_{\rm miss} \vert > 0.2 \gevc$ to suppress these events.
  We impose $m(\Dz \pi^\pm) - m(\Dz) > 0.16 \gevcc$ for the
  $\Dz \pip\pim \ell^- \nub$ channel
  to remove correctly reconstructed $\Bm \to \Dstarp \pim \ell^- \nub$ events
  with a subsequent $\Dstarp\to\Dz\pip$ decay.

  We use a separate Fisher discriminant~\cite{ref:Fisher} in each signal channel
  to further reduce the background from continuum and \BB\ events.
  The variables used are $E_{\rm extra}$, $m_{ES}$, 
  the number of unused neutral clusters with energy greater than $80 \mev$, the numbers of charged tracks and neutral clusters
  in the $B_{\rm tag}$ candidate, the second normalized Fox-Wolfram moment $R_2$~\cite{ref:FoxWolfram},
  and the CM-frame cosine of the angle 
  between the thrust 
  axes of the $B_{\rm tag}$ candidate and of the remaining particles in the event. 
  The discriminants are constructed using simulated events, with the distribution of each variable
  reweighted to match the distribution in data.
  The selection requirement on the
  output variables is optimized assuming a branching fraction
  $\mathcal{B}(\Bb \to D^{(*)} \pip \pim \ell^- \nub) = 0.12\%$ in each channel.

\begin{figure}[htb]
\begin{center}
\begin{picture}(230,150)
\put(0,0){
\begin{minipage}[b]{0.99\linewidth}
\includegraphics[width=0.9\textwidth , trim = 0.2cm 0.1cm 1.6cm 0.cm, clip]{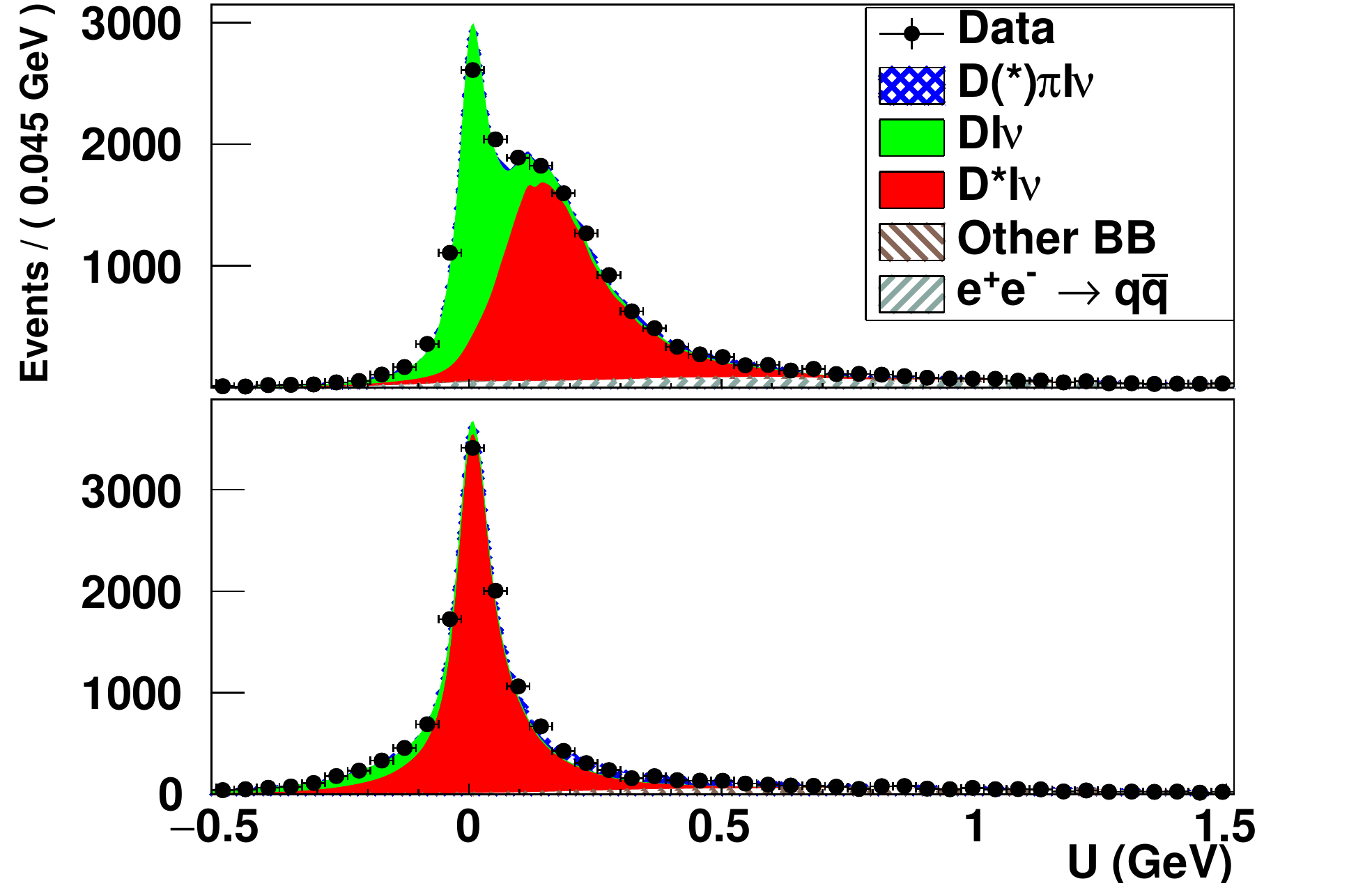}
\end{minipage}
}
\put(55,130){(a) \hspace{0.7cm}  $\Bm\to D^0\ell^-\nub$}
\put(108,120){\babar}
\put(55,65){(b)  \hspace{0.7cm} $\Bm\to D^{*0}\ell^-\nub$}
\put(108,55){\babar}
\end{picture}
\caption{Measured $U$ distributions and results of the fit for the
(a) $\Bm\to D^0\ell^-\nub$ and (b) $\Bm\to D^{*0}\ell^-\nub$ samples.
}
\label{fig:DzComposition}
\end{center}
\end{figure}

  At this stage of the analysis an event may be reconstructed in more
  than one channel.  To obtain statistically independent samples and to maximize the 
  sensitivity to $D^{(*)} \pip\pim \ell^- \nub$ decays, we select a unique candidate as follows.
  Any event
  found in a $D^{(*)} \ell^- \nub$ sample is removed from all samples 
  with one or two signal pions.  
  If an event enters two or more samples
  with the same number of signal pions, candidates are removed 
  from the sample with lower signal-to-background level. 
  In addition, we remove from the $D^{(*)} \pip\pim \ell^- \nub$ samples any
  event found in a $D^{(*)} \pi \ell^- \nub$ sample with $|U| < 0.1 \gev$.  
    
  The analysis procedure was developed using simulated event samples; the data for
  the two-pion signal modes were not examined until the selection and fit procedures were
  finalized.
  Event yields are obtained from an unbinned maximum likelihood fit to the $U$
  distribution in the range $-1.5 < U < 3.0 \gev$ for each signal channel. 
  One-dimensional probability density functions (PDF) for the signal and background 
  components of each sample are 
  obtained from 
  MC using parametric kernel estimators with 
  adaptive widths~\cite{ref:KeysPDF}.
  Figure~\ref{fig:DzComposition} shows the results for the 
  $D^{(*)0}\ell^-\nub$ channels; the results for the $D^{(*)+}\ell^-\nub$ channels are similar.
  Corresponding yields are presented in Table~\ref{tab:evtyieldsDPiPilnu}.
 
  The PDFs used in the fit to the $D^{(*)} \ell^-\nub$ channels 
  include the following components, whose magnitudes 
  are parameters of the fit: $\Bb \to D \ell^-\nub$, $\Bb \to \Dstar\ell^-\nub$, 
  $\Bb \to D^{(*)} \pi \ell^-\nub$, 
  other \BB\ events, and continuum events.  Potential contributions from $D^{(*)}\pi\pi\ell^-\nub$
  decays have a similar shape to $D^{(*)}\pi\ell^-\nub$ decays in these channels and are included
  in the $\Bb \to D^{(*)} \pi \ell^-\nub$ component. 
  The PDFs used in the fit to the $D^{(*)} \pip\pim\ell^-\nub$ channels include the following components:
  $\Bb \to D^{(*)} \ell^-\nub$,
  $\Bb \to D^{(*)} \pim \ell^-\nub$,
  $\Bb \to D \pip\pim\ell^-\nub$, $\Bb \to \Dstar \pip\pim\ell^-\nub$,
  other \BB\ events, and continuum events.
  Contributions to the $\Bb\to D^{(*)}\pip\pim\ell^-\nub$ channels
  from $\Bb\to D^{(*)}\pi^\pm\piz\ell^-\nub$ and $\Bb\to D^{(*)}\piz\piz\ell^-\nub$
  decays (cross-feed) are treated as signal.

  A fraction of signal decays are reconstructed with a $B$ meson charge differing by $\pm1$ from the 
  true $B$ meson charge and contribute 
  to the wrong signal channel.  We determine this fraction for each signal channel in simulation 
  and fix the corresponding yield ratio in the fit.
  Hadronic $B$ meson decays in which a hadron is misidentified as a lepton can peak near
  $U=0$.  We estimate these small contributions using simulation
  and hold them fixed in the fit to the $D^{(*)} \ell^-\nub$ channels.
  Simulation indicates that these peaking backgrounds 
  are negligible for the $D^{(*)}\pip\pim\ell^-\nub$ channels.

  Fits to ensembles of parameterized MC pseudo-experiments 
  are used to validate the fit.  All fitted parameters exhibit unbiased means and variances.

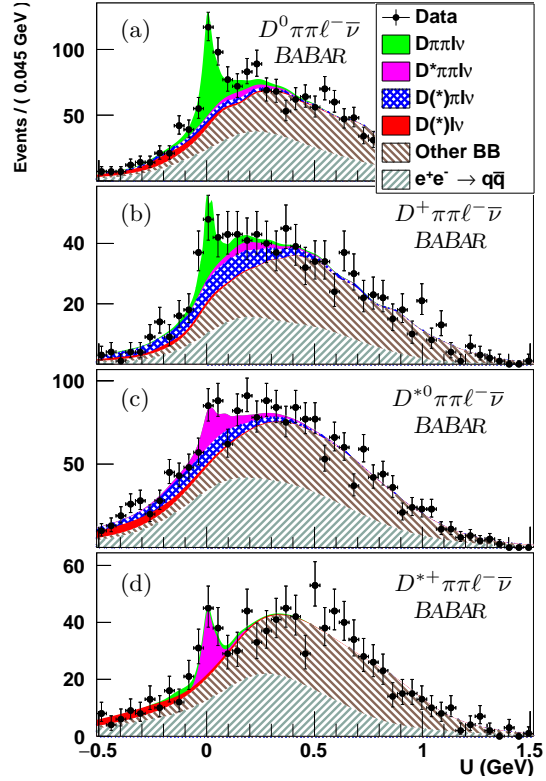
\begin{figure}[htb]
\begin{center}
\begin{picture}(230,300)
\put(0,0){
\begin{minipage}[b]{0.99\linewidth}
\includegraphics[width=0.9\textwidth , trim = 0.2cm 0.1cm 1.6cm 0.1cm, clip]{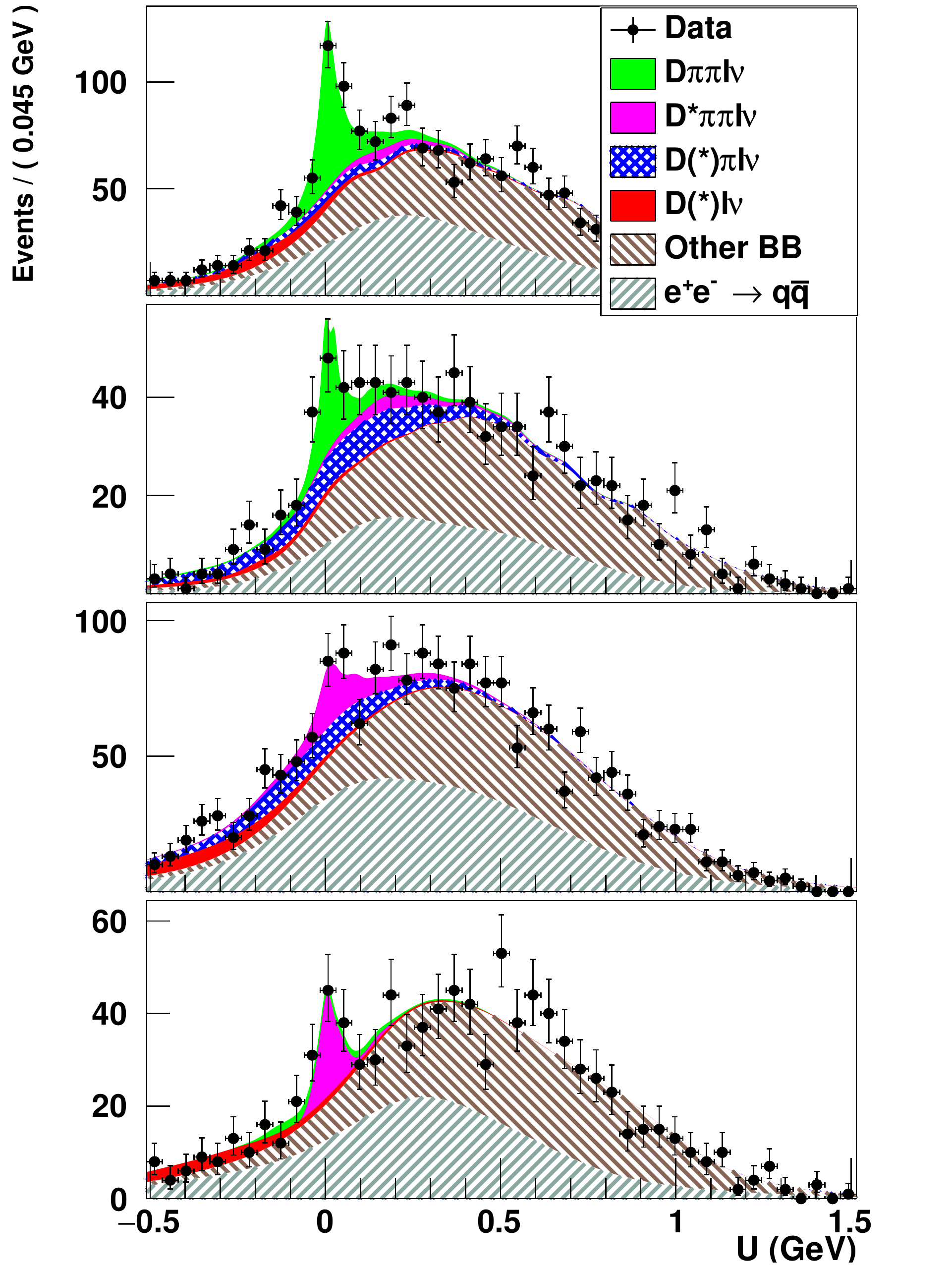} 
\end{minipage}
}
\put(55,280){(a) \hspace{1.2cm} $D^{0} \pi \pi \ell^-\nub$}
\put(113,270){\babar}
\put(55,210){(b) \hspace{3.0cm} $D^{+}\pi \pi  \ell^-\nub$}
\put(166,200){\babar}
\put(55,140){(c) \hspace{3.0cm} $D^{*0}\pi \pi  \ell^-\nub$}
\put(166,130){\babar}
\put(55,70){(d) \hspace{3.0cm} $D^{*+}\pi \pi   \ell^-\nub$}
\put(166,60){\babar}
\end{picture}
\caption{Measured $U$ distributions and results of the fit for the
  (a) $D^{0} \pi \pi \ell^-\nub$, (b) $D^{+}\pi \pi  \ell^-\nub$, 
  (c) $D^{*0}\pi \pi   \ell^-\nub$, and (d) $D^{*+}\pi \pi   \ell^-\nub$ samples.}
\label{fig:DPiPilnu}
\end{center}
\end{figure}

  The results for the $D^{(*)}\pip\pim\ell^-\nub$ channels are shown 
  in Fig.~\ref{fig:DPiPilnu} with the corresponding 
  signal yields in Table~\ref{tab:evtyieldsDPiPilnu}. 
  The fitted yields for all background components 
  are consistent with the values expected from MC. 
  The only known source of $\Bb\to D\pip\pim\ell^-\nub$ decays is $\Bb\to D_1(2420)\ell^-\nub$ with
  \mbox{$D_1(2420)\to D\pip\pim$}.  
  If we remove these $D_1(2420)$ decays by vetoing events with $0.5 < m(D\pip\pim)-m(D) < 0.6 \gevcc$, 
  the signal yields are reduced to 
  $84.3 \pm  27.7$ events in $\Dz\pip\pim$, and $37.3 \pm 15.9 $ in $\Dp\pip\pim$,
  which indicates that \mbox{$D_1(2420)\to D\pip\pim$} is not the only source for the observed signals.

\begin{table}[htb]
\caption{Event yields and estimated efficiencies ($\epsilon$) for the
 signal channels.
 The quoted uncertainties are statistical only. The fourth column gives the statistical significance, 
 $\mathcal{S} = \sqrt{2 \Delta \mathcal{L} }$, where $\Delta \mathcal{L}$ is the difference between 
 the log-likelihood value of the default fit and a fit with the signal yield fixed to zero.
 The last column gives the total significance, $\mathcal{S}_{tot}$,  where systematic uncertainties are included.}
\begin{center}
\begin{tabular}{|l|c|c|c|c|} \hline
Channel &  Yield  & $\epsilon \times 10^{4}$ & $ \mathcal{S}  $                 & $\mathcal{S}_{tot}$  \\ \hline\hline
$\Dz\ell^-\nulb $	     &	$5567 \pm 102$&	$2.73 \pm 0.01$  & $>40$ & $>40$ \\  
$\Dp\ell^-\nulb $	     &	$3236 \pm \,\ 74$&	$1.69 \pm 0.01$  & $>40$ & $>40$ \\  
$\Dstarz\ell^-\nulb $      &	$9987 \pm 126$&	$2.03 \pm 0.01$  & $>40$ & $>40$ \\  
$\Dstarp\ell^-\nulb $      &	$5404 \pm \,\ 83$&	$1.14 \pm 0.01$  & $>40$ & $>40$ \\  
$\Dz \pi\pi\ell^-\nub  $     &  $ 171 \pm 30$ & $1.18 \pm 0.03$ &  $5.4$      & $5.0$\\
$\Dp \pi\pi\ell^-\nub $      &  $\ 56 \pm 17$ & $0.51 \pm 0.02$ &  $3.5$      & $3.0$\\
$\Dstarz \pi\pi\ell^-\nub $ &   $\ 74 \pm 36$ & $1.11 \pm 0.02$ &  $1.8$      & $1.6$\\
$\Dstarp \pi\pi\ell^-\nub $ &   $\ 65 \pm 18$ & $0.49 \pm 0.02$ &  $3.3$      & $3.0$\\
\hline
\end{tabular}
\end{center}
\label{tab:evtyieldsDPiPilnu}
\end{table}

  Systematic uncertainties arising from limited knowledge of branching fractions, form factors, 
  and detector response are evaluated.  These impact the determination of the PDF shapes, fixed 
  backgrounds, cross-feed contributions, and signal
  efficiencies.  The leading uncertainties arise from
  ignorance of potential resonance structure in the $D^{(*)}\pip\pim$ final state,
  the limited size of MC samples 
  used to derive PDFs, 
  and the modeling of distributions of variables used in the Fisher discriminants.
  The dependence on the $D^{(*)}\pi\pi$ production process is investigated by using, in turn, 
  each of the individual mechanisms listed previously to model the signal.
  We assign the maximum deviation between the branching fraction ratios 
  $R^{(*)}_{\pip\pim}$ obtained from the nominal and alternative
  decay models as an uncertainty, giving
  $7.8\%$ for $\Dz\pip\pim\ell^-\nub$,
  $10.5\%$ for $\Dp\pip\pim\ell^-\nub$,
  $19.2\%$ for $\Dstarz\pip\pim\ell^-\nub$, and
  $13.4\%$ for $\Dstarp\pip\pim\ell^-\nub$.
  The impact of the statistical uncertainties of the PDFs are estimated from 
  fits to 1300 simulated data sets, obtained from the
  primary MC samples using the bootstrapping method~\cite{ref:bootstrapping},
  resulting in uncertainties
  ranging from $6.5\%$ ($\Dz\pip\pim\ell^-\nub$) to $21.1\%$ ($\Dstarz\pip\pim\ell^-\nub$).
  We estimate the uncertainty associated with modeling the Fisher discriminants
  by using the uncorrected shape of each simulated input distribution, one at a time, before imposing
  the selection requirement.  The systematic uncertainty, given by the sum in quadrature
  of the differences with respect to the nominal analysis, varies  
  from $3.7\%$ ($\Dz\pip\pim\ell^-\nub$) to $5.2\%$ ($\Dp\pip\pim\ell^-\nub$).

The ratios of branching fractions are calculated from the fitted yields as
\begin{equation}
R^{(*)}_{\pip\pim} = \frac{N_{\pip\pim}^{(*)}}{N_{\rm norm}^{(*)}}\frac{\epsilon_{\rm norm}^{(*)}}{\epsilon_{\pip\pim}^{(*)}},
\end{equation}
where $\epsilon$ refers to the corresponding efficiency, which is calculated from
MC for the same type of $B$ meson ($\Bm$ or $\Bzb$) used in the two-pion 
signal ($N_{\pip\pim}^{(*)}$) and zero-pion normalization ($N_{\rm norm}^{(*)}$) yields.  
The results are given in Table~\ref{tab:Rvalues}.
The dependence of the efficiencies
on the details of the hadronic $B$ reconstruction largely cancels in the ratio,
as do some other associated systematic uncertainties 
and possible biases.
Since semileptonic $B$ decays proceed via a spectator diagram, the semileptonic decay widths of neutral
and charged $B$ mesons are expected to be equal.  We therefore determine combined values for
the $\Bm$ and $\Bzb$ channels: these are given in Table~\ref{tab:Rvalues}.  Also shown are the
corresponding $\Bm$ branching fractions obtained by
using Ref.~\cite{ref:PDG2014} for the branching fractions of the normalization modes.

\begin{table}[!htb]
\caption{Branching fraction ratios $R^{(*)}_{\pip\pim}$ for the $D^{(*)}\pip\pim\ell^-\nub$ channels
and corresponding isospin-averaged values.
The first uncertainty is statistical and the second is systematic.  
The rightmost column gives the corresponding branching fractions, where the third uncertainty comes from the branching fraction of the normalization mode.  
The isospin-averaged results are quoted as $\Bm$ branching fractions.
}
\begin{center}
\begin{tabular}{|l|c|c|} \hline
Channel                  &        $R^{(*)}_{\pip\pim}\times 10^3$       & $\mathcal{B}\times 10^5$  \\[2pt] \hline\hline
$\Dz\pip\pim\ell^-\nub      $&  $71 \pm 13 \pm\,\ 8$           & $ 161 \pm 30 \pm 18 \pm 8$ \\    
$\Dp\pip\pim\ell^-\nub      $&  $58 \pm 18 \pm 12$             & $ 127 \pm 39 \pm 26 \pm 7$ \\   
$\Dstarz\pip\pim\ell^-\nub$&    $14 \pm\,\ 7  \pm\,\ 4$        & $ \,\ 80 \pm 40 \pm 23 \pm 3$ \\  
$\Dstarp\pip\pim\ell^-\nub$&    $28 \pm\,\ 8  \pm\,\ 6$        & $ 138 \pm 39 \pm 30 \pm 3$ \\ \hline  
$D\pip\pim\ell^-\nub      $&    $67 \pm 10 \pm\,\ 8$           & $ 152 \pm 23 \pm 18 \pm 7$ \\  
$\Dstar\pip\pim\ell^-\nub  $&   $19 \pm\,\ 5 \pm\,\ 4$         & $ 108 \pm 28 \pm 23 \pm 4$ \\  
\hline
\end{tabular}
\end{center}
\label{tab:Rvalues}
\end{table}

In conclusion, 
the decays $\Bb\to D^{(*)} (n \pi) \ell^-\nub$ with $n=0$ or $2$ are studied in events with
a fully reconstructed
  second $B$ meson.  We obtain the first observation of
$\Bb\to \Dz\pip\pim\ell^-\nub$ decays and first evidence for
$\Bb\to D^{(*)+}\pip\pim\ell^-\nub$ decays.  The branching
ratios of $\Bb\to D^{(*)}\pip\pim\ell^-\nub$ decays relative to the corresponding 
$\Bb\to D^{(*)}\ell^-\nub$ decays are measured.
To estimate the total $\Bb\to D^{(*)}\pi\pi\ell^-\nub$ branching fraction we use
isospin symmetry and consider in turn each of the $\Bb\to X_c\ell^-\nub$
decay models discussed above.
We find
$\mathcal{B}(\Bb\to D^{(*)}\pip\pim\ell^-\nub)/\mathcal{B}(\Bb\to D^{(*)}\pi\pi\ell^-\nub)=0.50\pm 0.17$, where the uncertainty is one half the observed spread from the investigated models, 
which implies
$\mathcal{B}(\Bb\to D\pi\pi\ell^-\nub)+\mathcal{B}(\Bb\to D^{*}\pi\pi\ell^-\nub)=(0.52 {}^{+0.14}_{-0.07} {}^{+0.27}_{-0.13})\%$, where the first 
uncertainty is the total experimental uncertainty and the second is due to the unknown fraction of 
$\Bb\to D^{(*)}\pip\pim\ell^-\nub$ in $\Bb\to D^{(*)}\pi\pi\ell^-\nub$ decays.
This corresponds to between one-quarter and one-half of the difference between the sum of the
previously measured exclusive $B$ meson semileptonic decays to charm final states and the corresponding
inclusive semileptonic branching fraction.

We are grateful for the excellent luminosity and machine conditions
provided by our \pep2\ colleagues, 
and for the substantial dedicated effort from
the computing organizations that support \babar.
The collaborating institutions wish to thank 
SLAC for its support and kind hospitality. 
This work is supported by
DOE
and NSF (USA),
NSERC (Canada),
CEA and
CNRS-IN2P3
(France),
BMBF and DFG
(Germany),
INFN (Italy),
FOM (The Netherlands),
NFR (Norway),
MES (Russia),
MINECO (Spain),
STFC (United Kingdom),
BSF (USA-Israel). 
Individuals have received support from the
Marie Curie EIF (European Union)
and the A.~P.~Sloan Foundation (USA).


\end{document}